\documentstyle[epsf,twoside,fleqn,espcrc2]{article}

\newcommand{\AmS}{{\protect\the\textfont2
  A\kern-.1667em\lower.5ex\hbox{M}\kern-.125emS}}

\title{Spin Models on Thin Graphs}

\author{C. F. Baillie\address{Computer Science Department,\\
        Campus Box 430, University of Colorado,\\ Boulder CO80309, USA}%
        and 
        D. A. Johnston\address{Mathematics Department, Heriot-Watt University, \\
        Edinburgh, EH14 4AS, United Kingdom}}
       
\begin{document}

\begin{abstract}
We discuss the utility of analytical and numerical investigation
of spin models, in particular spin glasses, on ordinary ``thin'' random graphs
(in effect Feynman diagrams) using 
methods borrowed from the ``fat'' graphs of two dimensional gravity.
We highlight the similarity with Bethe lattice calculations
and the advantages of the thin graph approach both analytically
and numerically for investigating mean field results.
\end{abstract}

\maketitle

\section{INTRODUCTION}

The analytical
investigation of spin glasses on random graphs of various sorts has a long
and honourable history \cite{1,1a},
though there has been little in the
way of numerical simulations. 
Random graphs with a fixed or fixed
average connectivity have a locally tree like structure, which means that
loops in the graph are predominantly large, so Bethe-lattice-like \cite{3}
(ie mean field) critical behaviour is expected for spin models on such lattices.
Given this, the analytical solution for a spin model
or, in particular, a spin glass on a Bethe lattice \cite{4,4a}
can be translated across to the appropriate
fixed connectivity random lattice.
Alternatively, a replica calculation can be carried out directly
in some cases
for spin glasses on various sorts of random lattices.

A rather different way of looking at the problem of spin models on random graphs was put forward in
\cite{5}, where it was observed that the requisite ensemble of random graphs
could be generated by considering
the Feynman diagram expansion for
the partition function of the model.
For an Ising ferromagnet with Hamiltonian
\begin{equation}
H = \beta \sum_{<ij>} \sigma_i \sigma_j,
\end{equation}
where the sum is over nearest neighbours on three-regular random graphs
(ie $\phi^3$ Feynman diagrams),
the partition function is given by
\begin{displaymath}
Z_n(\beta)  N_n = {1 \over 2 \pi i} \oint { d \lambda \over
\lambda^{2n + 1}} \int {d \phi_+ d \phi_- \over 2 \pi \sqrt{\det K}}
\exp (- S ) 
\end{displaymath}
where $N_n$ is the number of undecorated graphs with $2n$ points,
$K$ is defined by
\begin{equation}
\begin{array}{cc} K_{ab} = & \left(\begin{array}{cc}
\sqrt{g} & { 1 \over \sqrt{g}} \\
{1 \over \sqrt{g}} & \sqrt{g}
\end{array} \right) \end{array}
\end{equation}
and the action itself is
\begin{equation}
S = {1 \over 2 } \sum_{a,b}  \phi_a  K^{-1}_{ab} \phi_b  -
{1 \over 3} (\phi_+^3 + \phi_-^3).
\label{e3}
\end{equation}
where the sum runs over $\pm$ indices.
The coupling in the above is $g = \exp ( 2 \beta J )$
where $J=1$ for the ferromagnet and
the $\phi_+$ field can be thought of as representing ``up'' spins
with the $\phi_-$ field representing ``down'' spins.
An ensemble of $z$-regular random graphs would
simply require replacing the $\phi^3$ terms with $\phi^z$ and a fixed
average connectivity could also be implemented with the appropriate
choice of potential.

This approach was inspired by the considerable amount of work that has been done
in recent years on $N \times N$ matrix \footnote{$N$, the size
of the matrix
is not to be confused with $n$, the number of vertices in the graph!}
versions of such integrals which generate
``fat'' or ribbon graphs graphs
with sufficient structure to carry out a topological expansion \cite{6}
because of the matrix index structure.
The natural interpretation of such fat graphs as the duals
of triangulations, quadrangulations etc. of surfaces has led to much interesting work in string
theory and particle physics \cite{7}.
The partition function here is a poor, ``thin''
(no indices, so no ribbons), scalar cousin of these,
lacking the structure to give a surface interpretation to the graph. Such scalar
integrals have been used in the past
to extract the large $n$ behaviour of various
field theories \cite{8} again essentially as
a means of generating the appropriate Feynman diagrams, so a lot is known about handling their quirks.

\section{(ANTI)FERROMAGNETS AND SPIN GLASSES}

For the Ising ferromagnet on three-regular ($\phi^3$) graphs,
solving the saddle point equations at large $n$
\begin{eqnarray}
\phi_+ &=&  \sqrt{g} \phi_+^2 + {1 \over \sqrt{g}} \phi_-^2 \nonumber \\
\phi_- &=&  \sqrt{g} \phi_-^2 + {1 \over \sqrt{g}} \phi_+^2
\end{eqnarray}
shows that the critical behaviour appears as an exchange of dominant saddle point solutions
to the saddle point equations.
The high and low temperature solutions respectively are
\begin{eqnarray}
\phi_+,\phi_- &=&  {  \sqrt{g} \over  g + 1
         }  \nonumber \\
\phi_+,\phi_- &=& { \sqrt{g} \over 2 ( g - 1)} \left( 1 \pm   \sqrt{ { g  - 3 \over g + 1} } \right)
\end{eqnarray}
which give a low temperature magnetized phase.
The critical exponents for the transition can also be calculated
in this formalism and, as expected, are mean field. In general
a mean field transition appears at
\begin{equation}
\exp ( 2 \beta_{FM}) = z / ( z - 2)
\end{equation}
on $\phi^z$ graphs,
which is the value predicted by
the standard approaches.
Simulations nicely confirm this mean field picture for the ferromagnet
\cite{pap1}. 
Analysis of the Binder's cumulant for the magnetization also
shows that the critical temperatures are identical to
the corresponding Bethe lattices (ie $g=3$ for $\phi^3$ graphs). 
The specific heat is shown
in Figure.1 for various sizes of $\phi^3$ graphs.

\par
\par
\vspace{1.0in}
\begin{figure}[h]
\vspace{2.0in}
\includegraphics{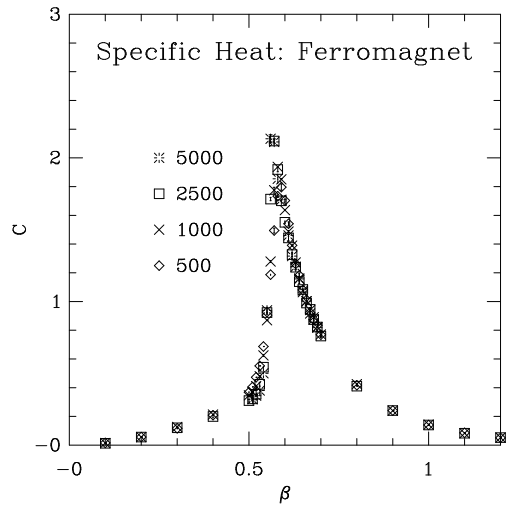}
\vspace{0.5in} 
\caption{The specific Heat for the Ising ferromagnet on $\phi^3$
graphs of various size.}
\label{fig1:}
\end{figure}

There are various possibilities for addressing
spin glass order in the Feynman diagram approach.
In \cite{5} the entropy per spin was calculated
for the Ising {\it anti}-ferromagnet on $\phi^3$ graphs and
it was found to become negative for sufficiently negative
$\beta$, which is often indicative of a spin glass transition.
Simulations again confirm the picture. Taking a quenched distribution
of couplings of the form 
\begin{equation}
P(J) = p \; \delta ( J -1)
+ ( 1 - p) \; \delta ( J + 1 ),
\end{equation}
which gives the antiferromagnet for $p=0$, produces
results for the spin glass order parameter, the overlap,
that are very similar to the infinite range mean field
(Sherrington-Kirkpatrick) model.
Defining the overlap as
\begin{equation}
q = {1 \over n} \sum_i \sigma_i \tau_i.
\end{equation}
with two Ising replicas on each graph, $\sigma_i, \tau_i$
and histogramming 
\begin{equation}
P_n(q) = \left[ \langle \delta ( q - 1/ n \sum \sigma_i \tau_i) \rangle \right],
\end{equation}
where $[ \; ]$ denotes the quenched disorder average,
we get the distribution shown in Figure.2 \cite{pap1,pap2} in the putative 
spin glass phase at low 
temperature. The long tail stretching down to
$q=0$ is characteristic of the mean-field spin glass picture
of many inequivalent states.

\begin{figure}[h]
\vspace{2.0in}
\includegraphics{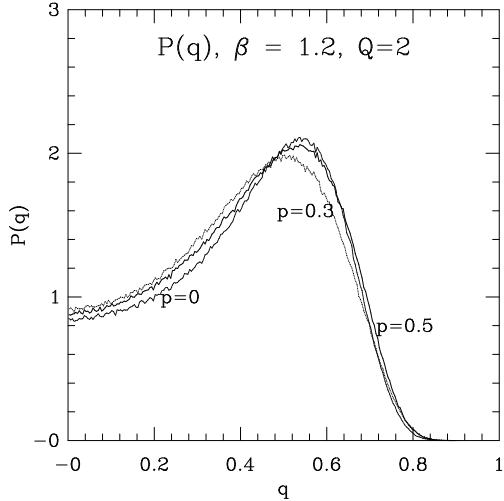}
\vspace{0.5in}
\caption{$P(q)$ at $\beta=1.2$ for various $p$ on $\phi^3$ graphs of size 500.}
\label{fig2:}
\end{figure}

It is possible to make some analytical inroads as well
by looking at the solutions to the saddle point equations for
$k$ Ising replicas \cite{pap2}
\begin{equation}
\vec \phi = \ \int \;  P (J) \otimes^k K \; dJ
 \; \vec \phi^{2} 
\end{equation}
where we have denoted the $2^k$ fields that now appear as $\vec \phi$.
The Hessian for these equations is analytically calculable for 
{\it any} $k$
\begin{equation}
\prod_{m=0}^{k} \left[  2 \int P(J) \tanh (\beta J)^m dJ - 1 \right]^{{k} \choose {m}}
\end{equation}
and its zeroes show that the $k=0$ transition 
temperature observed in simulations or calculated by using the analogy
with the Bethe lattice
is identical to the $k=2$ transition temperature.
This also occurs in the finite replica version of the
Sherrington-Kirkpatrick model \cite{sk2}, so yet again the
thin graph results are resolutely mean field.

For three or more replicas one does
not see the continuous transition
because a first order transition occurs at higher temperature
to a replica-symmetric state. The situation appears to be rather
similar for $Q>2$ state Potts models where the saddle point calculation
finds a continuous transition at one of the spinodal points and misses
a first order transition occurring at higher temperature.
The 3-state Potts model, for instance, with action
\begin{eqnarray}
S &=& { 1 \over 2 } ( \phi_a^2 + \phi_b^2 + \phi_c^2 ) - c ( \phi_a \phi_b + \phi_a \phi_c + \phi_b \phi_c) \nonumber \\
&-& {1 \over 3} ( \phi_a^3 + \phi_b^3 + \phi_c^3 ),
\label{potts3}
\end{eqnarray}
gives high and low temperature solutions
\begin{eqnarray}
\phi_{a,b,c} &=& 1 - 2 c \; ; \nonumber \\
\nonumber \\
\phi_{a,b} &=& { 1 + \sqrt{1 - 4 c - 4 c^2} \over 2} \; ,\nonumber \\
\phi_c &=& { 1 + 2 c - \sqrt{1 - 4 c - 4 c^2} \over 2}
\label{potts3sol}
\end{eqnarray}
where $c = 1/ ( g + 1)$.

\begin{figure}[ht]
\vspace{2.0in}
\includegraphics{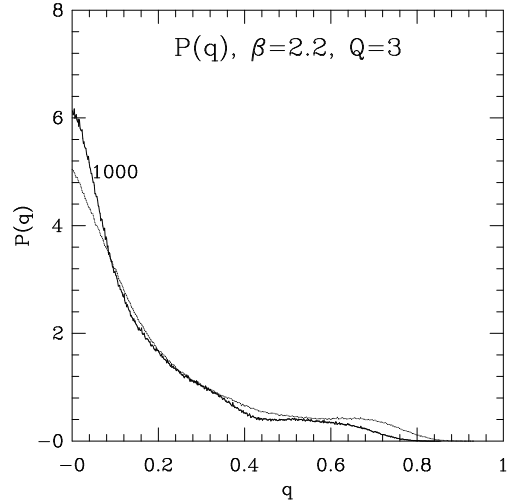}
\vspace{0.5in}
\caption{$P(q)$ at $\beta=2.2$ for the 3-state Potts model for $\phi^3$ graphs
of size 1000 (labelled) and 500 (unlabelled).}
\label{fig3:}
\end{figure}

Calculations and simulations for Potts glasses are just as easy as for the
Ising spin glass.
In Figure.3 overleaf we have plotted the distribution of overlaps
for the three state Potts model at low temperature. 
The overlap for a Q state Potts model is now defined as
\begin{equation}
q = { 1 \over n} \sum_{i=1}^n ( Q  \delta_{\sigma_i , \tau_i} - 1 )
\end{equation}
and the lack of an inversion symmetry in the spins gives
a different pattern of replica symmetry breaking in mean-field
theory to the Ising model. 
The numerical results are still
consistent with a mean field picture.

\section{Conclusions}

In summary, spin models on thin graphs offer a promising arena
for the application of ideas from matrix models, large-$n$
calculations in field theory
and bifurcation theory. 
In the spin glass case the tensor (or near-tensor) product structure
of the inverse propagator allows some quite general expressions
to be derived for the Hessian in the saddle point equations
and offers a powerful line of attack on questions such as
replica symmetry breaking. As a subject for numerical simulations
they offer the great advantage of mean field results with no
infinite range interactions and no boundary problems.

The bulk of the simulations were carried out
on the Front Range Consortium's
208-node Intel Paragon located at NOAA/FSL in Boulder.
CFB is supported by DOE under
contract DE-FG02-91ER40672, by NSF Grand Challenge Applications
Group Grant ASC-9217394 and by NASA HPCC Group Grant NAG5-2218.
CFB and DAJ were partially supported by NATO grant CRG910091.


\begin{thebibliography}{99}
\bibitem{1} L. Viana and A. Bray, J. Phys. {\bf C 18} (1985) 3037.
\bibitem{1a} M. Mezard and G. Parisi, Europhys. Lett. {\bf 3} (1987) 1067;\\
            I. Kanter and H. Sompolinsky, Phys. Rev. Lett. {\bf 58} (1987) 164;\\
            K Wong and D. Sherrington, J. Phys. {\bf A21} (1988) L459;\\
            C. de Dominicis and Y. Goldschmidt, Phys. Rev. {\bf B41} (1990) 2184;\\            P-Y Lai and Y. Goldschmidt, J. Phys. {\bf A23} (1990) 399.
\bibitem{3} H. A. Bethe, Proc. Roy. Soc. {\bf A 150} (1935) 552;\\
            T. P. Eggarter, Phys. Rev. {\bf B9} (1974) 2989;\\
            E. Muller-Hartmann and J. Zittartz, Phys. Rev. Lett. {\bf
            33} (1974) 893.
\bibitem{4} S. Inawashiro and S. Katsura, Physica {\bf 100A} (1980) 24;\\
             D. Thouless, Phys. Rev. Lett. {\bf 56} (1986) 1082;\\
             P. Mottishaw, Europhys. Lett. {\bf 4} (1987) 333;\\
             K Wong and D. Sherrington, J. Phys. {\bf A20} (1987) L785.
\bibitem{4a} Y. Goldschmidt, Europhys. Lett. {\bf 6} (1988) 7.
\bibitem{5} C. Bachas, C. de Calan and P. Petropoulos, J. Phys. {\bf A27}
            (1994) 6121.
\bibitem{6} E. Brezin, C. Itzykson, G. Parisi and J.B. Zuber,
            Commun. Math. Phys. {\bf 59} (1978) 35;\\
            M.L. Mehta, Commun. Math. Phys. {\bf 79} (1981) 327.
\bibitem{7} For a review see, J. Ambjorn, `` Quantization of Geometry''
            Les Houches 1994, hep-th/9411179.
\bibitem{8} J. Le Guillou and J. Zinn-Justin (editors), ``Large Order Behaviour
            of Perturbation Theory'', Amsterdam: North Holland  (1989).
\bibitem{pap1} C.F. Baillie, D. A. Johnston and J-P. Kownacki, Nucl. Phys. {\bf B432} (1994) 551.
\bibitem{pap2} C.F. Baillie, W. Janke, D. A. Johnston and P. Plechac, ``Spin Glasses
on Thin Graphs'', to appear in NPB[FS].
\bibitem{sk2}D. Sherrington, J. Phys. {\bf A13} 637 (1980);\\
             R. Penney, A. Coolen and D. Sherrington, J. Phys. {\bf A26}
             3681 (1993).
\end{thebibliography}
\end{document}